\documentclass[a4paper]{spie}  %>>> use this instead for A4 paper
\usepackage{amsmath,amsfonts,amssymb}
\usepackage{graphicx}
\usepackage[colorlinks=true, allcolors=blue]{hyperref}

\usepackage{array}
\usepackage{units}
\usepackage{textcomp}
\usepackage{amsmath}
\usepackage{amssymb}
\usepackage{esint}
\usepackage{amsfonts}
\usepackage{multirow}
\usepackage{rotating}
\usepackage{multicol}
\usepackage{hyperref}
\usepackage{subcaption}
\usepackage{wrapfig}
\usepackage{sidecap}
\usepackage{color}
\usepackage{caption}
\usepackage{paralist}
\usepackage{authblk}

\newcommand{\be}{\begin{equation}}
\newcommand{\ee}{\end{equation}}
\def \ea{e-ASTROGAM }
\def \eap{e-ASTROGAM}

\def\gsim{\lower.5ex\hbox{$\; \buildrel > \over \sim \;$}}
\title{The e-ASTROGAM gamma-ray space observatory for the multimessenger astronomy of the 2030s}

\author[a]{V. Tatischeff}
\author[b,c,d,e]{A. De Angelis}
\author[f,g,h]{M. Tavani}
\author[i]{I. Grenier}
\author[j]{U. Oberlack}
\author[k]{L. Hanlon}
\author[l]{R.~Walter}
\author[m]{A. Argan}
\author[n]{P. von Ballmoos}
\author[o]{A. Bulgarelli}
\author[f]{I. Donnarumma}
\author[p]{M. Hernanz}
\author[q]{I.~Kuvvetli}
\author[b]{M. Mallamaci}
\author[r]{M. Pearce}
\author[s]{A. Zdziarski}
\author[t,c]{A. Aboudan}
\author[u]{M. Ajello} 
\author[v]{G.~Ambrosi}		
\author[w]{D. Bernard}
\author[x]{E. Bernardini} 	
\author[y]{V. Bonvicini}
\author[j]{A. Brogna}
\author[h,aa]{M. Branchesi}	
\author[q]{C.~Budtz-Jorgensen}
\author[aa]{A. Bykov}
\author[o]{R. Campana}
\author[f]{M. Cardillo}
\author[v]{S. Ciprini}
\author[ab]{P. Coppi}
\author[a]{P.~Cumani}
\author[ac]{R.M. Curado da Silva}
\author[ad]{D. De Martino}	
\author[ae,af]{R. Diehl}
\author[b,t]{M. Doro}
\author[o]{V. Fioretti}
\author[ag]{S. Funk}
\author[ah]{G. Ghisellini}
\author[ai,aj]{F. Giordano}
\author[ak]{J.E. Grove}
\author[a]{C. Hamadache}
\author[u]{D.H.~Hartmann}
\author[al]{M. Hayashida}		
\author[p]{J. Isern}
\author[am]{G.~Kanbach}
\author[a]{J.~Kiener}
\author[n]{J. Kn\"odlseder}
\author[o]{C. Labanti}
\author[an]{P. Laurent}
\author[u]{M. Leising}
\author[ao]{O. Limousin}
\author[ap,y]{F.~Longo}
\author[aq]{K.~Mannheim}
\author[ar]{M. Marisaldi}
\author[as]{M. Martinez} 
\author[aj]{M.N.~Mazziotta}  	
\author[at]{J.E.~McEnery}
\author[au]{S.~Mereghetti}
\author[f]{G. Minervini}
\author[av]{A. Moiseev}
\author[aw]{A. Morselli}
\author[ax]{K. Nakazawa}
\author[ay]{P.~Orleanski} 
\author[az]{J.M. Paredes}		
\author[ba]{B. Patricelli}		
\author[a]{J. Peyr\'e}
\author[f]{G. Piano}
\author[bb]{M. Pohl}
\author[b,t]{R. Rando}
\author[bc]{M. Roncadelli} 		
\author[ah]{F. Tavecchio}	 
\author[at]{D. J. Thompson}
\author[t]{R. Turolla}
\author[k]{A. Ulyanov}
\author[d,y]{A.~Vacchi}
\author[bd]{X. Wu}
\author[be]{A. Zoglauer}

\affil[a]{CSNSM, CNRS/Univ. Paris-Sud, Universit\'e Paris-Saclay, F-91405 Orsay Campus, France}
\affil[b]{INFN, Sezione di Padova, I-35131 Padova, Italy}
\affil[c]{INAF - Osservatorio Astronomico di Padova, I-35122, Padova, Italy}
\affil[d]{Dipartimento di Matematica, Informatica e Fisica, Universita di Udine, I-33100 Udine, Italy}
\affil[e]{Laboratorio de Instrumentacao e Particulas and Instituto Superior Tecnico, Lisboa, Portugal}
\affil[f]{INAF-IAPS, via del Fosso del Cavaliere 100, I-00133 Roma, Italy}
\affil[g]{Dip. di Fisica, Univ. Roma Tor Vergata, via della Ricerca Scientifica 1, I-00133 Roma, Italy}
\affil[h]{Gran Sasso Science Institute, viale Francesco Crispi 7, I-67100 L'Aquila, Italy}
\affil[i]{AIM Paris-Saclay, CEA/IRFU, CNRS, Univ Paris Diderot, 91191 Gif-sur-Yvette, France}
\affil[j]{Institute of Physics and PRISMA Excellence Cluster, Johannes Gutenberg University Mainz, D-55099 Mainz, Germany}
\affil[k]{School of Physics, University College Dublin, Ireland}
\affil[l]{University of Geneva, Chemin d’Ecogia 16, CH-1290 Versoix, Switzerland}
\affil[m]{INAF Headquarters, Viale del Parco Mellini, 84, I-00136, Roma, Italy}
\affil[n]{IRAP Toulouse, 9 av. du Colonel-Roche - BP 44 346, F-31028 Toulouse Cedex 4, France}
\affil[o]{INAF/IASF Bologna, Via Gobetti 101, I-40129 Bologna, Italy}
\affil[p]{ICE, CSIC and IEEC, Campus UAB, Carrer Can Magrans s/n, E-08193 Cerdanyola del Valles, Barcelona,
Spain}
\affil[q]{DTU Space, National Space Institute, Technical University of Denmark, Kgs. Lyngby, Denmark}
\affil[r]{KTH Royal Institute of Technology, Dept. of Physics, 10691 Stockholm, Sweden}
\affil[s]{Nicolaus Copernicus Astronomical Center, Polish Academy of Sciences, Bartycka 18, PL-00-716
Warszawa, Poland}
\affil[t]{Dipartimento di Fisica e Astronomia ``G. Galilei'', Universit\`a di Padova, I-35131 Padova, Italy}
\affil[u]{Dept. of Physics and Astronomy, Clemson University, Clemson, SC 29634, USA}
\affil[v]{INFN, Sezione di Perugia, I-06123, Perugia, Italy}
\affil[w]{LLR, Ecole Polytechnique, CNRS/IN2P3, F-91128 Palaiseau, France}
\affil[x]{Deutsches Elektronen Synchrotron (DESY), Platanenallee 6, D-15738, Zeuthen, Germany}
\affil[y]{INFN, Sezione di Trieste, via A. Valerio, I-34127 Trieste, Italy}
\affil[z]{INFN, Laboratori Nazionali del Gran Sasso, L'Aquila, Italy}
\affil[aa]{Ioffe Institute, St.Petersburg 194021, Russia}
\affil[ab]{Dept. of Astronomy, Yale University, P.O. Box 208101, New Haven, CT 06520-8101, USA}
\affil[ac]{LIP, Departamento de F´ısica Universidade de Coimbra, P-3004-516 Coimbra, Portugal}
\affil[ad]{INAF - Osservatorio Astronomico di Capodimonte, Salita Moiariello 16, I-80131 Napoli, Italy}
\affil[ae]{Max Planck Institut fuer Extraterrestrische Physik, Giessenbachstr.1, D-85748 Garching, Germany}
\affil[af]{Excellence Cluster Universe, Germany}
\affil[ag]{Friedrich-Alexander-Universit¨at Erlangen-N¨unberg, Erwin-Rommel-Str. 1, D-91058 Erlangen,
Germany}
\affil[ah]{INAF - Osservatorio di Brera, via E. Bianchi 46, I-23807 Merate, Italy}
\affil[ai]{Dipartimento di Fisica dell'Universit\`a di Bari, Bari, Italy}
\affil[aj]{INFN Sezione di Bari, I-70126 Bari, Italy}
\affil[ak]{U.S. Naval Research Laboratory, 4555 Overlook Ave SW, Washington, DC 20375, USA}
\affil[al]{Institute for Cosmic Ray Research, the University of Tokyo, Kashiwa, Chiba, 277-8582, Japan}
\affil[am]{Max-Planck-Institut fur Extraterrestrische Physik, Postfach 1312, D-85741 Garching, Germany}
\affil[an]{APC, Univ Paris Diderot, CNRS/IN2P3, CEA/Irfu, Observatoire de Paris, 10 rue Alice Domont et L´eonie Duquet, F-75205 Paris Cedex 13, France}
\affil[ao]{CEA/Saclay IRFU/Department of Astrophysics, Bat. 709, F-91191, Gif-Sur-Yvette, France}
\affil[ap]{Dipartimento di Fisica, Universita di Trieste, I-34127 Trieste, Italy}
\affil[aq]{Universitaet Wuerzburg, Lehrstuhl fuer Astronomie, D-97074 Wuerzburg, Germany}
\affil[ar]{University of Bergen, Norway}
\affil[as]{IFAE-BIST, Edifici Cn. Universitat Autonoma de Barcelona, E-08193 Bellaterra, Spain}
\affil[at]{NASA Goddard Space Flight Center, Greenbelt, MD, USA}
\affil[au]{INAF/IASF, Via Bassini 15, I-20133 Milano, Italy}
\affil[av]{CRESST/NASA/GSFC and University of Maryland, College Park, USA}
\affil[aw]{INFN Roma Tor Vergata , Via della Ricerca Scientifica 1,  I-00133 Roma, Italy}
\affil[ax]{Department of Physics, The University of Tokyo, 7-3-1 Hongo, Bunkyo-ku, Tokyo 113-0033}
\affil[ay]{Space Research Center of Polish Academy of Sciences, Bartycka 18a, PL-00-716 Warszawa, Poland}
\affil[az]{Departament de F\'isica Qu\`antica i Astrof\'isica, ICCUB, Universitat de Barcelona, IEEC-UB, Mart\'i i Franqu\`es 1, E-08028 Barcelona, Spain}
\affil[ba]{Scuola Normale Superiore, Piazza dei Cavalieri 7, I-56126 Pisa, and INFN Pisa, Italy}
\affil[bb]{Institute of Physics and Astronomy, University of Potsdam, 14476 Potsdam, Germany}
\affil[bc]{INFN Pavia, via A. Bassi 6, I-27100 Pavia, Italy; INAF Milano, Milano, Italy}
\affil[bd]{DPNC, 24 Quai Ernest-Ansermet, CH-1211 Gen\`eve 4, Switzerland}
\affil[be]{University of California, Space Sciences Laboratory, 7 Gauss Way, Berkeley, CA 94720, USA}

\authorinfo{On behalf of the e-ASTROGAM Collaboration (see \url{http://eastrogam.iaps.inaf.it}). \\
Corresponding author e-mail addresses: vincent.tatischeff@csnsm.in2p3.fr; alessandro.deangelis@pd.infn.it
}

\begin{document} 
\maketitle

\begin{abstract}
\ea is a concept for a breakthrough observatory space mission carrying a $\gamma$-ray telescope dedicated to the study of the non-thermal Universe in the photon energy range from 0.15 MeV to 3 GeV. The lower energy limit can be pushed down to energies as low as 30 keV for gamma-ray burst detection with the calorimeter. The mission is based on an advanced space-proven detector technology, with unprecedented sensitivity, angular and energy resolution, combined with remarkable polarimetric capability. 
Thanks to its performance in the MeV--GeV domain,  substantially improving its predecessors, \ea will open a new window on the non-thermal 
Universe, making pioneering observations of the most powerful  Galactic and extragalactic  sources, elucidating the nature of their relativistic outflows and  their effects on the surroundings. 
With a line sensitivity in the MeV energy range one to two orders of magnitude better than previous and current generation instruments, \ea will determine the origin of key isotopes fundamental for the understanding of supernova explosion and the chemical evolution of our Galaxy. The mission will be a major player of the multiwavelength, multimessenger time-domain astronomy of the 2030s, and provide unique  data of significant interest to a broad astronomical community, complementary to powerful observatories such as LISA, LIGO, Virgo, KAGRA, the Einstein Telescope and the Cosmic Explorer, IceCube-Gen2 and KM3NeT, SKA, ALMA, JWST, E-ELT, LSST, Athena, and the Cherenkov Telescope Array. 

\end{abstract}

\keywords{Gamma-ray astronomy, time-domain astronomy, space mission, Compton and pair creation telescope, gamma-ray polarization, high-energy astrophysical phenomena}

\section{INTRODUCTION}
\label{sec:intro}  % \label{} allows reference to this section

\ea \cite{eas} is a concept for a breakthrough gamma-ray space observatory operating in a maturing multimessenger epoch, opening up entirely new and exciting synergies. The mission will provide unique and complementary gamma-ray data of significant interest to a broad astronomical community, in a decade of powerful observatories for multiwavelength astronomy and for the detection of gravitational waves, neutrinos and ultra-high-energy cosmic rays (UHECRs). 

The core mission science of \ea addresses three major topics of modern astrophysics.

\begin{figure*}[t]
\centering
\includegraphics[width=0.6\textwidth]{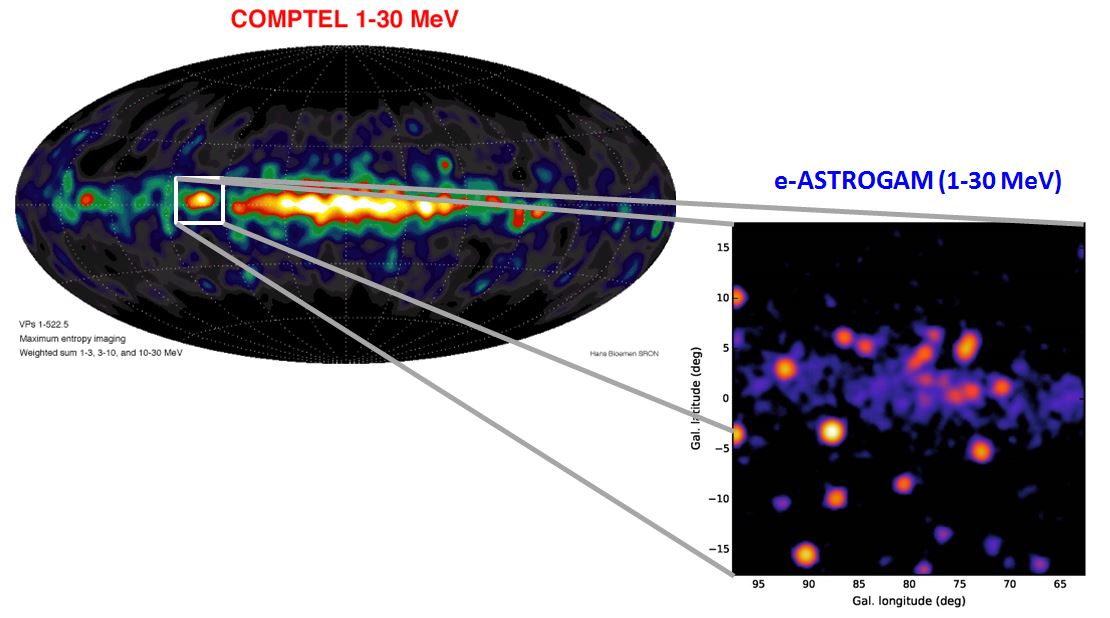}
\includegraphics[width=0.675\textwidth]{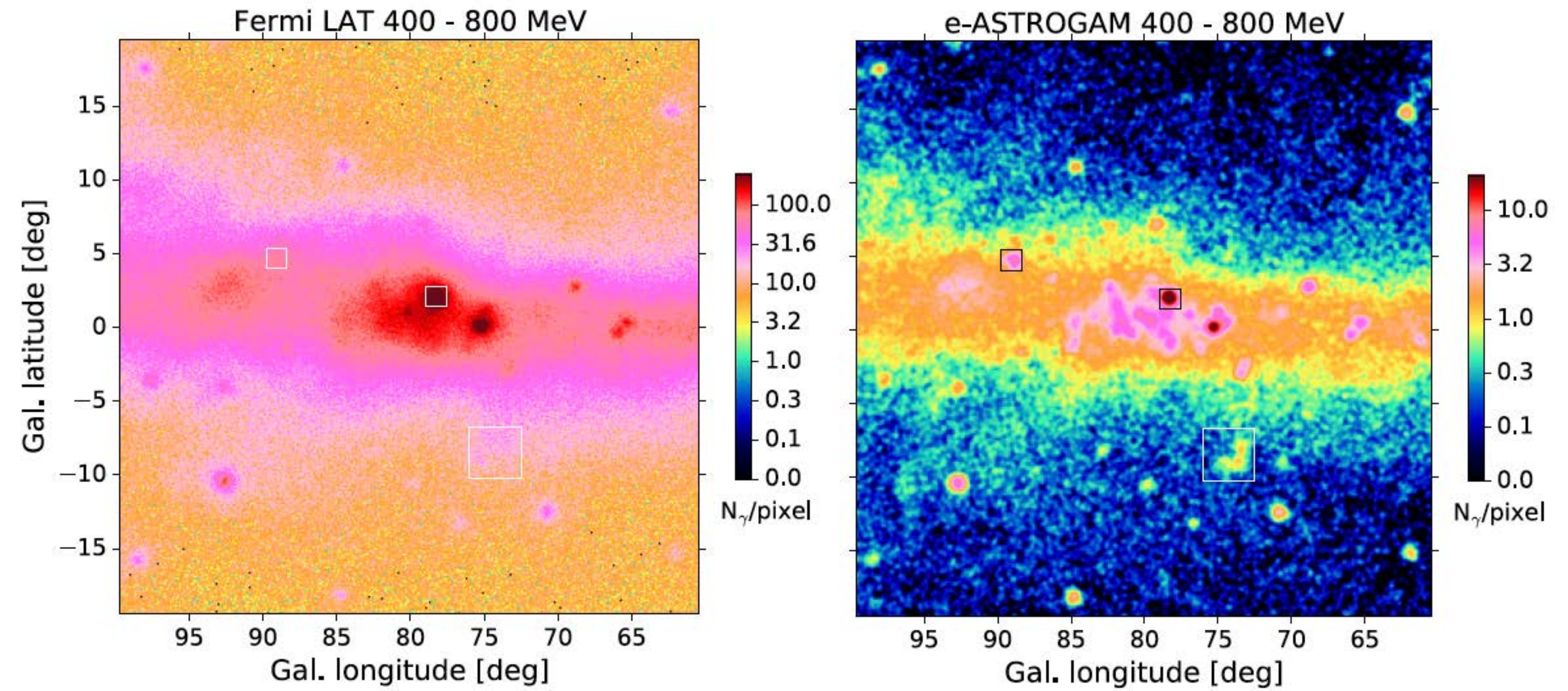}
\vspace{0.3cm}
\caption{An example of the capability of \ea to transform our 
 knowledge of the MeV-GeV sky. Upper panel: the upper left figure shows the 1-30 MeV sky as observed by COMPTEL in the 1990s and the upper right figure shows the simulated Cygnus region in the 1-30 MeV
energy region  from \eap. Lower panel: comparison between the view of the Cygnus region by Fermi in 8 years (left) and that by e-ASTROGAM in one year of effective exposure (right) between 400 and 800 MeV.}
\label{fig:Jurgen}
\end{figure*}

\begin{itemize} \itemsep 0cm \topsep 0cm
\item \textbf{\em{Processes at the heart of the extreme Universe: prospects for the astronomy of the 2030s}}

Observations of relativistic jet and outflow sources (both in our Galaxy and in active galactic nuclei, AGNs) in the X-ray and  GeV--TeV energy ranges have shown that the  MeV--GeV band holds the key to understanding the  transition from the low energy continuum to a spectral range shaped by very poorly understood particle acceleration processes. In many cases, a substantial fraction of the radiated power appears in the MeV band, and the unprecedented sensitivity of \ea in this energy domain (see Fig.~\ref{fig:sensitivity}) will offer an exceptional view of the violent processes operating close to supermassive black holes (BHs), inside the powerful explosions that we see as gamma-ray bursts (GRBs), and during the merger of two neutron stars (NSs) or of a NS and a BH. \ea will: (1) determine the composition (hadronic or leptonic) of the outflows and jets, which strongly influences the environment -- breakthrough polarimetric capability and spectroscopy 
providing the keys to unlocking this long-standing question;
(2) identify the physical acceleration processes in these outflows and jets (e.g. diffusive shocks, magnetic field reconnection, plasma effects), that may lead to dramatically different particle energy distributions; (3) clarify the role of the magnetic field in powering ultrarelativistic jets in GRBs, through time-resolved polarimetry and spectroscopy.

In addition, measurements in the \ea energy band will have a big  impact on multimessenger astronomy in the 2030s. Joint gravitational-wave and gamma-ray observations of high-energy transients are key to obtain a more complete knowledge of the sources and their environments, since these two closely-related messengers provide complementary information. \ea observations in the MeV--GeV band can also play a decisive role for our understanding of the astrophysical sources of high-energy neutrinos and UHECRs.  

\begin{figure*}[t]
\centering
\includegraphics[width=0.85\textwidth]{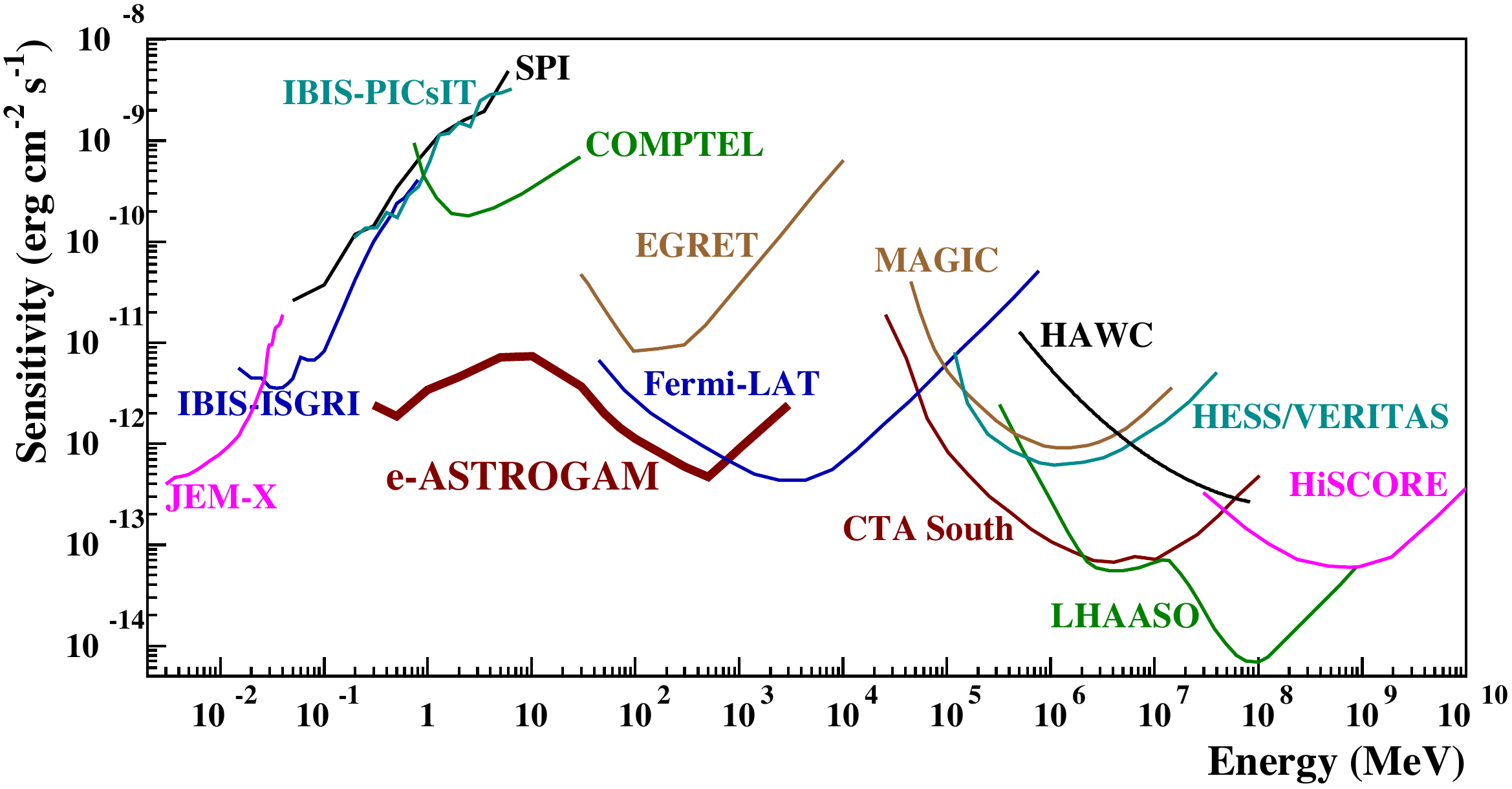}
\vspace{0.2cm}
\caption{{Point source continuum differential sensitivity of different X- and $\gamma$-ray instruments. The curves for {\it INTEGRAL}/JEM-X, IBIS (ISGRI and PICsIT), and SPI are for an effective observation time $T_{\rm obs}$ = 1 Ms. The COMPTEL and EGRET sensitivities are given for {the typical observation time accumulated during the $\sim$9 years of the {\it CGRO} mission (see Fig. 1 in Ref.\cite{tak13}). The {\it Fermi}/LAT sensitivity is for a high Galactic latitude source in 10 years of  observation in survey mode}. For MAGIC, VERITAS (sensitivity of H.E.S.S. is similar), and CTA, the sensitivities are given for $T_{\rm obs}$ = 50 hours. For HAWC $T_{\rm obs}$ = 5 yr, for LHAASO $T_{\rm obs}$ = 1~yr, and for HiSCORE $T_{\rm obs}$ = 1000 h. The e-ASTROGAM sensitivity is calculated at $3\sigma$ for an effective exposure of 1 year and for a source at high Galactic latitude.}
\label{fig:sensitivity}}
\end{figure*}

\item \textit{\textbf{The origin and impact of cosmic-ray particles on galaxy evolution}}

\ea  will resolve the outstanding issue of the origin and propagation of low-energy cosmic rays in the interstellar medium (ISM) of the Milky Way. It  will measure cosmic-ray diffusion in interstellar clouds and their impact on gas dynamics and state, thus providing crucial diagnostics about the cosmic-ray feedback on star formation, ISM structures, galactic winds and outflows, as well as on the growth of interstellar magnetic field. The improved sensitivity and angular resolution of \ea (see Fig.~\ref{fig:Jurgen}) will be crucial to probe the interplay between cosmic rays and the turbulent medium of star forming regions (e.g., Cygnus Cocoon) during the early steps of their Galactic voyage. Sensitive \ea observations of a set of cosmic-ray sources, such as young supernova remnants, will allow for the first time to distinguish the emission produced by the interactions of cosmic-ray nuclei with the ambient gas and the non-thermal emission from cosmic-ray electrons. Combined with high-resolution radio and X-ray observations, the \ea data will provide information on cosmic-ray injection into the acceleration process, on the structure of magnetic fields inside the remnants, and on the spectrum of cosmic rays freshly released into surrounding clouds.

\item \textbf{\textit{Nucleosynthesis and the chemical enrichment
of our Galaxy}}

The \ea gamma-ray line sensitivity is more than an order of magnitude  better than that of previous or current instruments.  The deep exposure of the Galactic plane region will determine how different isotopes ($^{26}$Al, $^{44}$Ti, $^{60}$Fe) are created in stars and distributed in the interstellar medium; it will also unveil the recent history of supernova explosions in the Milky Way and detect the radioactive emission of Galactic classical novae ($^{7}$Be, $^{22}$Na, 511~keV line) for the first time. \ea will also detect a significant number of supernovae in nearby galaxies, thus addressing fundamental issues in the explosion mechanisms of both core-collapse and thermonuclear supernovae. The gamma-ray data ($^{56}$Ni, $^{56}$Co) will provide a much better understanding of Type Ia supernovae and their evolution with look-back time and metallicity, which is a pre-requisite for their use as standard candles for precision cosmology.  

\end{itemize}

In addition to addressing its core scientific goals, \ea would 
achieve many serendipitous discoveries through its combination of wide field of view and improved sensitivity, measuring  in three years the spectral energy distributions of thousands of Galactic and extragalactic sources. An extensive review of the science achievable with this mission is given in the recently submitted \ea White Book \cite{wb}. \ea can become a key contributor to multiwavelength time-domain astronomy as an Observatory facility open to a wide astronomical community. 

Table \ref{tab:nevents} summarizes our conservative estimates of the number of sources detectable by e-ASTROGAM in three years, based on current knowledge and $\log N - \log S$ determinations of Galactic and extragalactic sources, including GRBs. It takes information from the {the \textit{Swift}-BAT 70-Month Hard X-ray survey catalog \cite{b70h}, the 4th \textit{INTEGRAL}-IBIS catalog \cite{ibisc}, and the 3rd \textit{Fermi}-LAT catalog \cite{3FGL}. Noteworthy, the latter catalog contains more than 1000 unidentified sources in the 100 MeV -- 300 GeV range with no counterparts at other wavelength\footnote{The Preliminary LAT 8-year Point Source List (FL8Y) contains 2132 unidentified sources, see \url{https://fermi.gsfc.nasa.gov/ssc/data/access/lat/fl8y/}}, and most of them will be detected by e-ASTROGAM, in addition to  a relevant number of new unidentified sources. The discovery space of e-ASTROGAM for new sources and source classes is very large.}

\vskip 2mm
\begin{table}
\begin{center}
\begin{tabular}{| l | l | l |}\hline
Type & 3 yr & New sources\\ \hline
%All sky (above 100 MeV) & $> 3000$ & $\sim$1800 (including GRBs)  \\
{Total} & {3000 -- 4000} & $\sim$1900 (including GRBs)  \\
Galactic & $\sim1000$ & $\sim$400 \\
%Galactic sources $(>$ 30 MeV) & &\\
MeV blazars  & $\sim350$ & $\sim350$ \\
GeV blazars  & {1000 -- 1500} & $\sim350$ \\
Other AGN  ($< $10 MeV)& {70 -- 100} & {35 -- 50}\\
Supernovae  & {10 -- 15} & {10 -- 15}\\
Classical novae & 4 -- 6 &   4 -- 6 \\
Long GRBs  & $\sim$540 &  $\sim$540\\
Short GRBs  & $\sim$180 &  $\sim$180\\ \hline
\end{tabular}
\end{center}
\caption{Estimated number of sources {of various classes} detectable by e-ASTROGAM in three years. The last column gives the number of sources not known before in any  wavelength. \label{tab:nevents}}
\end{table}

An important characteristic of \ea is its ability to measure  polarization in the MeV range, which is afforded by Compton interactions in the detector. Polarization encodes information about the geometry of magnetic fields and adds a new observational pillar, in addition to the temporal and spectral, through which fundamental processes governing the MeV emission can be determined. The addition of polarimetric information will be crucial for a variety of investigations, including accreting BH systems, magnetic field structures in jets, and the emission mechanisms of GRBs. Polarization will also provide definitive insight into the presence of hadrons in extragalactic jets and the origin of UHECRs.

The e-ASTROGAM mission concept is presented at length in Ref.~\cite{eas}. Here, we first give an overview of the proposed observatory (Sect.~\ref{sect:ea}), then discuss the complementarity of \ea to gravitationnal wave and neutrino astronomies (Sect.~\ref{sect:multimessenger}), and finally outline the breakthrough capability of the e-ASTROGAM telescope for gamma-ray polarimetric observations (Sect.~\ref{sect:pol}; see also Ref.~\cite{ea_jatis}).

\section{The e-ASTROGAM observatory}
\label{sect:ea} 

\begin{figure}[t]
\centering
\includegraphics[width=0.75\textwidth]{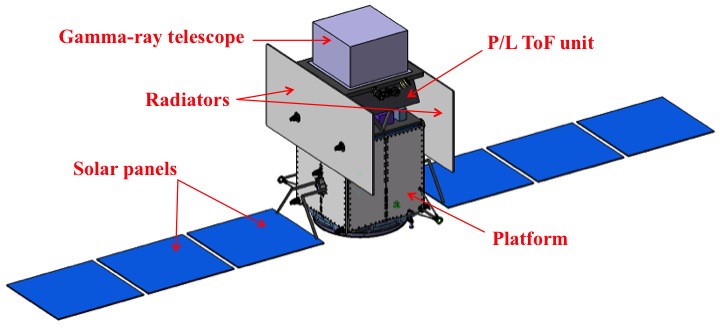}
\caption{e-ASTROGAM spacecraft with solar panels deployed.}
\label{fig:deployed}
\end{figure}

The payload of the e-ASTROGAM satellite (Figure~\ref{fig:deployed}) consists of a single gamma-ray telescope operating over more than four orders of magnitude in energy (from about 150 keV to 3 GeV) by the joint detection of photons in both the Compton (0.15 -- 30 MeV) and pair ($> 10$~MeV) energy ranges. It is attached to a mechanical structure at a distance of about 90~cm from the top of the spacecraft platform, the space between the payload and the platform being used to: (i) host a time-of-flight (ToF) unit designed to discriminate between particles coming out from the telescope and those entering the instrument from below; (ii) host several units of the payload (the back-end electronics modules, the data handling unit, and the power supply unit) and (iii) accommodate two fixed radiators of the thermal control system, each of 5.8~m$^2$ area (Figure~\ref{fig:deployed}). This design has the advantage of significantly reducing the instrument background due to prompt and delayed gamma-ray emissions from fast particle reactions with the platform materials. 

\begin{figure}[t]
\begin{center}
\begin{tabular}{c}
\begin{minipage}{0.42\linewidth}
\includegraphics[scale=0.3]{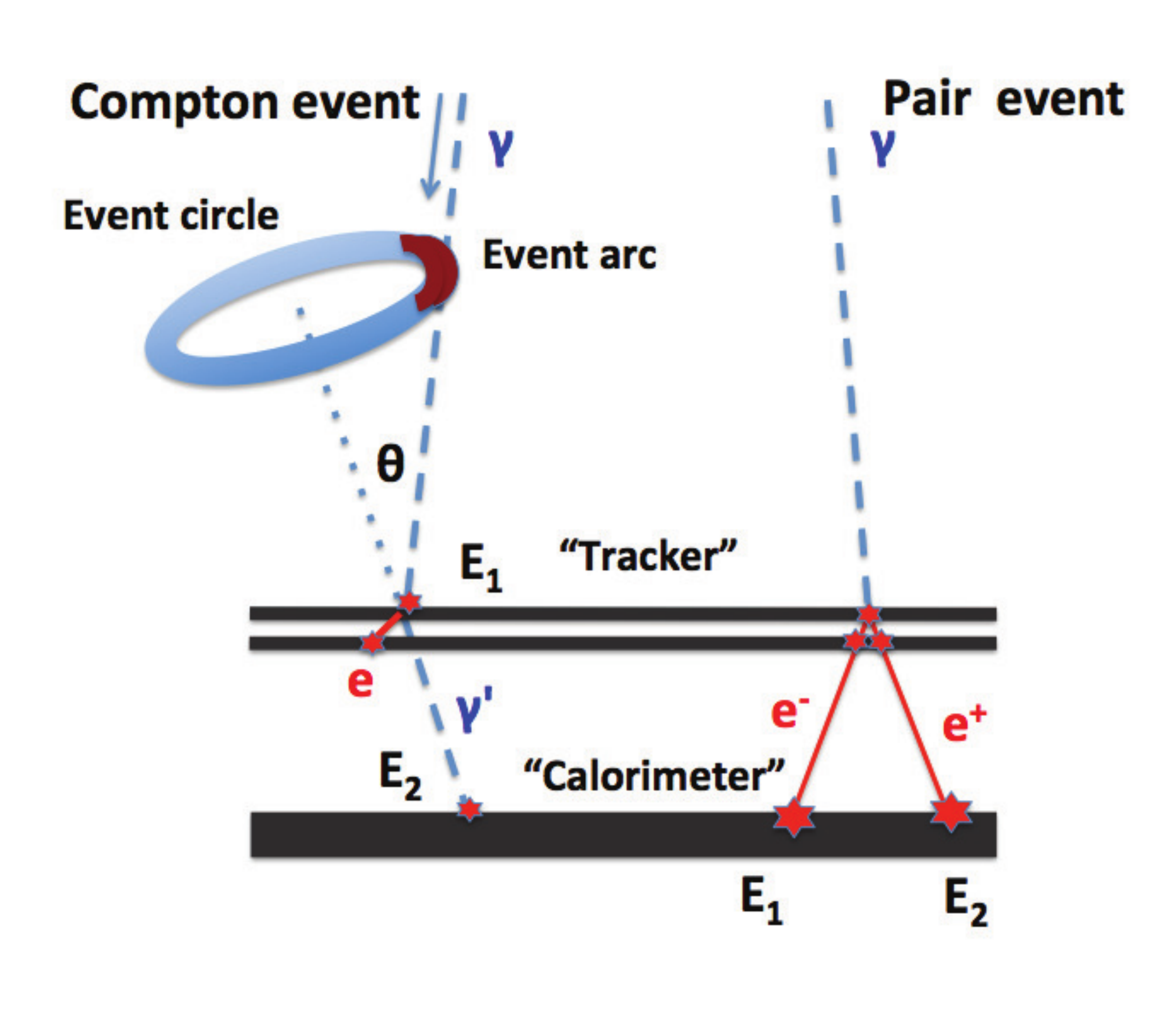}
\end{minipage}
\begin{minipage}{0.58\linewidth}
\includegraphics[scale=0.22]{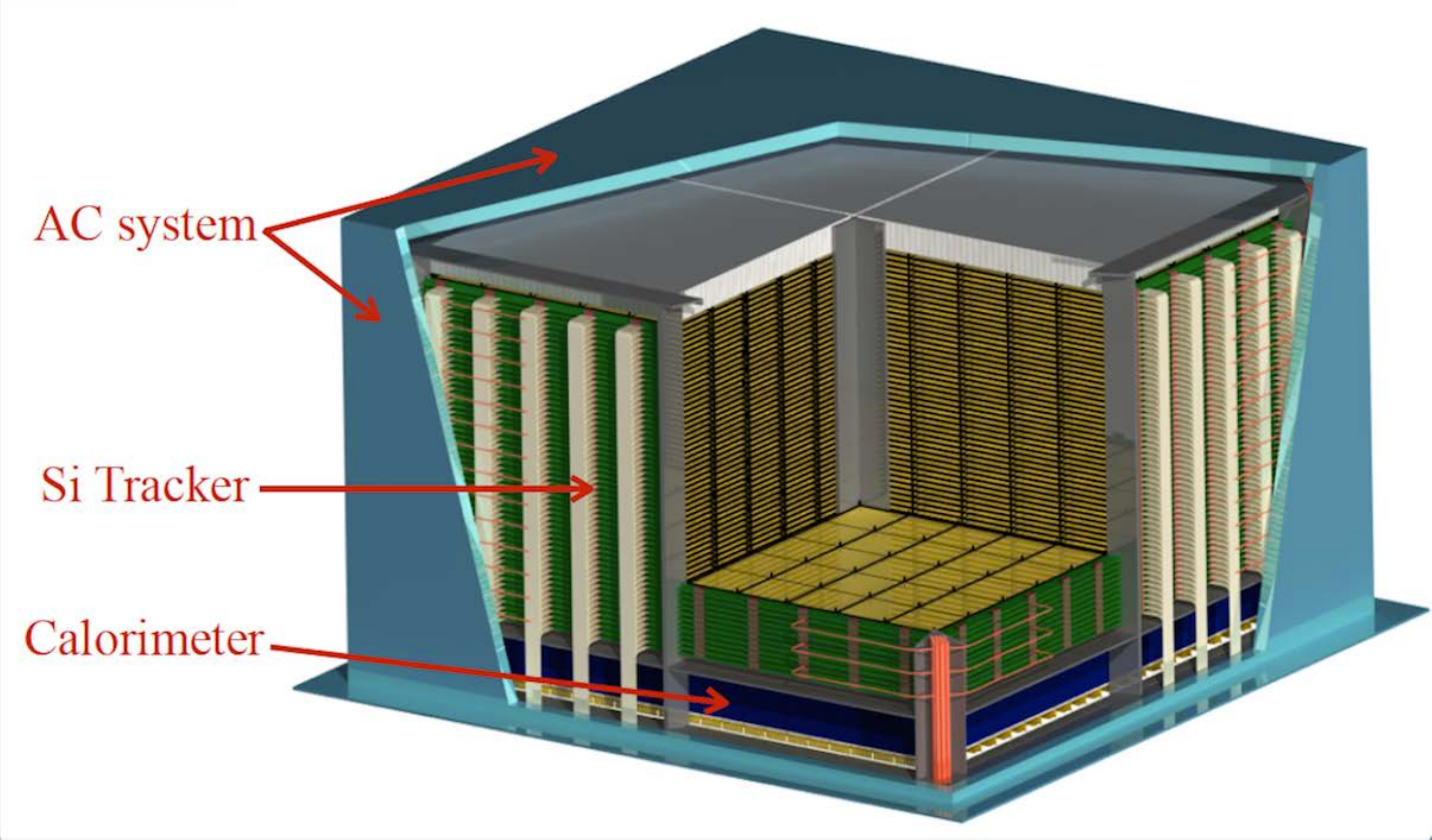}
\end{minipage}
\\
(a) \hspace{8.2cm} (b)
\end{tabular}
\end{center}
\caption 
{ \label{fig:payload}
(a) Representative  topologies for a Compton event and  for a pair event. Photon tracks are shown in pale blue, dashed, and electron and/or positron tracks are in red, solid.  (b) Overview of the e-ASTROGAM payload.} 
\end{figure}

The e-ASTROGAM telescope is made up of three detection systems (Figure~\ref{fig:payload}): a silicon Tracker in which the cosmic gamma-rays undergo a Compton scattering or a pair conversion (see Figure~\ref{fig:payload}a); a Calorimeter to absorb and measure the energy of the secondary particles and an anticoincidence (AC) system to veto the prompt-reaction background induced by charged particles. The telescope has a size of 120$\times$120$\times$78 cm$^3$ and a mass of 1.2~tons (including maturity margins plus an additional margin of 20\% at system level).

The Si Tracker comprises 5600 double-sided strip detectors (DSSDs) arranged in 56 layers. It is divided in four units of 5$\times$5 DSSDs, the detectors being wire bonded strip to strip to form 2-D ladders. Each DSSD has a geometric area of 9.5$\times$9.5 cm$^2$, a thickness of 500~$\mu$m, and a strip pitch of 240~$\mu$m. The total detection area amounts to 9025 cm$^2$. Such a stacking of relatively thin detectors enables efficient tracking of the electrons and positrons produced by pair conversion, and of the recoil electrons produced by Compton scattering. The DSSD signals are read out by 860,160 independent, ultra low-noise and low-power electronics channels with self-triggering capability.

The Calorimeter is a pixelated detector made of a high-$Z$ scintillation material -- Thallium activated Cesium Iodide -- for efficient absorption of Compton scattered gamma-rays and electron-positron pairs. It consists of an array of 33,856 parallelepiped bars of CsI(Tl) of 8~cm length and 5$\times$5~mm$^2$ cross section, read out by silicon drift detectors (SDDs) at both ends, arranged in an array of 529 ($=23 \times 23$) elementary modules each containing 64 crystals. The depth of interaction in each crystal is measured from the difference of recorded scintillation signals at both ends. Accurately measuring the 3D position and deposited energy of each interaction is essential for a proper reconstruction of the Compton events. The Calorimeter thickness -- 8 cm of CsI(Tl) -- makes it a 4.3 radiation-length detector having an absorption probability of a 1-MeV photon on-axis of 88\%.

The third main detector of the e-ASTROGAM payload consists of an Anticoincidence system composed of two main parts: (1) a standard Anticoincidence, named Upper-AC, made of segmented panels of plastic scintillators covering the top and four lateral sides of the instrument, requiring a total active area of about 5.2~m$^2$, and (2) a Time of Flight (ToF) system, aimed at rejecting the particle background produced by the platform. The Upper-AC detector is segmented in 33 plastic tiles (6 tiles per lateral side and 9 tiles for the top) coupled to silicon photomultipliers (SiPM) by optical fibers. The bottom side of the instrument is protected by the ToF unit, which is composed of two plastic scintillator layers separated by 50 cm, read out by SiPMs connected to Time Digital Converters. The required timing resolution is 300 ps (1$\sigma$). 

For best environmental conditions, the e-ASTROGAM satellite should be launched into a quasi-equatorial (inclination $i < 2.5^\circ$) low-Earth orbit (LEO) at a typical altitude of 550~--~600~km. The background environment in such an orbit is now well-known thanks to the Beppo-SAX\cite{cam14} and {\it AGILE}\cite{tav09} missions. In addition, such a LEO is practically unaffected by precipitating particles originating from solar flares, a virtue for background rejection. 

Extensive simulations of the instrument performance using state-of-art numerical tools\cite{zog06,bul12} and a detailed numerical mass model of the satellite together with a thorough model for the background environment have shown that e-ASTROGAM will achieve: 
\begin{itemize}
\item Broad energy coverage ($\sim$0.15 MeV to 3 GeV), with nearly two orders of magnitude improvement of the continuum sensitivity in the range 0.15 -- 100 MeV compared to previous missions (Fig.~\ref{fig:sensitivity};
\item Excellent sensitivity for the detection of key gamma-ray lines e.g. sensitivity for the 847~keV line from thermonuclear supernovae 70 times better than that of the {\it INTEGRAL} spectrometer (SPI);
\item Unprecedented angular resolution both in the MeV domain and above a few hundreds of MeV  i.e. improving the angular resolution of the COMPTEL telescope on board the {\it Compton Gamma Ray Observatory} ({\it CGRO}) and that of the {\it Fermi}/LAT instrument by a factor of $\sim$4 at 5 MeV and 1~GeV, respectively (e.g. the e-ASTROGAM Point Spread Function (68\% containment radius) at 1 GeV is 9').
\item Large field of view ($>$ 2.5 sr), ideal to detect transient Galactic and extragalactic sources, such as X-ray binaries and GRBs;
\item Timing accuracy of 1~$\mu$s (at 3$\sigma$), ideal to study the physics of magnetars and rotation-powered pulsars, as well as the properties of terrestrial gamma-ray flashes;
\item Pioneering polarimetric capability for both steady and transient sources. 
\end{itemize}

\section{e-ASTROGAM and the multimessenger astronomy}
\label{sect:multimessenger}  % \label{} allows reference to this section

\ea is uniquely positioned to be the space observatory of the 2030s with most complementarity for gravitational wave (GW) and neutrino astronomy. 

\subsection{e-ASTROGAM and gravitational wave astronomy}

e-ASTROGAM's very large field of view (more than 1/5 of the sky for the Tracker and even larger for the Calorimeter in transient mode starting from 30 keV) will be essential to properly cover the GW error boxes and find electromagnetic (EM) counterparts to the GW events. The recent binary NS merger generating the GW event GW170817 and the corresponding short GRB detected by Fermi GBM and INTEGRAL 1.7~s after the GW signal \cite{abb17} has shown that the soft gamma-ray domain is the most appropriate band of the EM spectrum to identify the source and define the astrophysical context of the burst event. In contrast, the number of optical transients spatially and temporally coincident with GW events is expected to be of the order of hundreds \cite{nis13}. 

The {\it on-axis} prompt emission of GRBs associated to GW events shall be favorably detected with e-ASTROGAM. When Advanced LIGO and Advanced Virgo will operate at design sensitivity, the expected range for GW detection is 200 Mpc for NS--NS mergers and 1 Gpc for BH--NS systems \cite{abb16}. From the most recent NS--NS merger rate estimates \cite{abb17}, the expected detection rate of the GRB on-axis prompt emission by e-ASTROGAM in coincidence with a GW detection is between 0.6~yr$^{-1}$ and 9~yr$^{-1}$; these numbers will double after the incorporation of KAGRA and LIGO-India into the GW network. \ea will be able to quickly transmit spectral and positional information to the community to trigger the GRB follow-up by both ground-based and space facilities. In addition to serendipitous observations, e-ASTROGAM will be able to effectively point at sky regions with GW identifications, the mission requirement for the satellite repointing  in case of a Target of Opportunity (ToO) observation being within 6--12 hours, with the goal of reaching 3--6 hours. 

\ea simulations based on the large GRB database \cite{batse_grbc,swift_grbc,gbm_grbc} yield detection rates of about 60 short GRBs and 180 long GRBs per year in the ``Gamma-ray imager'' trigger mode of the gamma-ray telescope (Table~\ref{tab:nevents}). Additional, softer bursts will be detected by the ``Calorimeter burst search'' mode of data acquisition (i.e. using triggers generated only by an increase of the Calorimeter count rate). The excellent sensitivity of \ea in the soft gamma-ray domain will also be crucial to reveal nearby {\it off-axis} GRBs associated to GW events. The 6$\sigma$ trigger threshold in the ``Calorimeter burst search'' mode is $\sim 0.05$~ph~cm$^{-2}$~s$^{-1}$ in the 100--300~keV energy range over 1~s timescale, which is an order of magnitude lower than the measured flux in the main pulse ($\Delta t =0.576$~s) of GRB~170817A ($F_\gamma$(100--300~keV)$=0.49$~ph~cm$^{-2}$~s$^{-1}$ according to the {\it {\it Fermi}}/GBM best-fit, Comptonized model, for this burst \cite{gol17}). GRB170817A is likely to have been observed at an angle of $\sim 30^\circ$ from the jet axis \cite{kim17}, and such an off-axis burst would be detected with \ea up to a distance of about $130$~Mpc. 

\begin{figure}[t]
\begin{center}
\begin{tabular}{c}
\begin{minipage}{0.47\linewidth}
\includegraphics[scale=0.525]{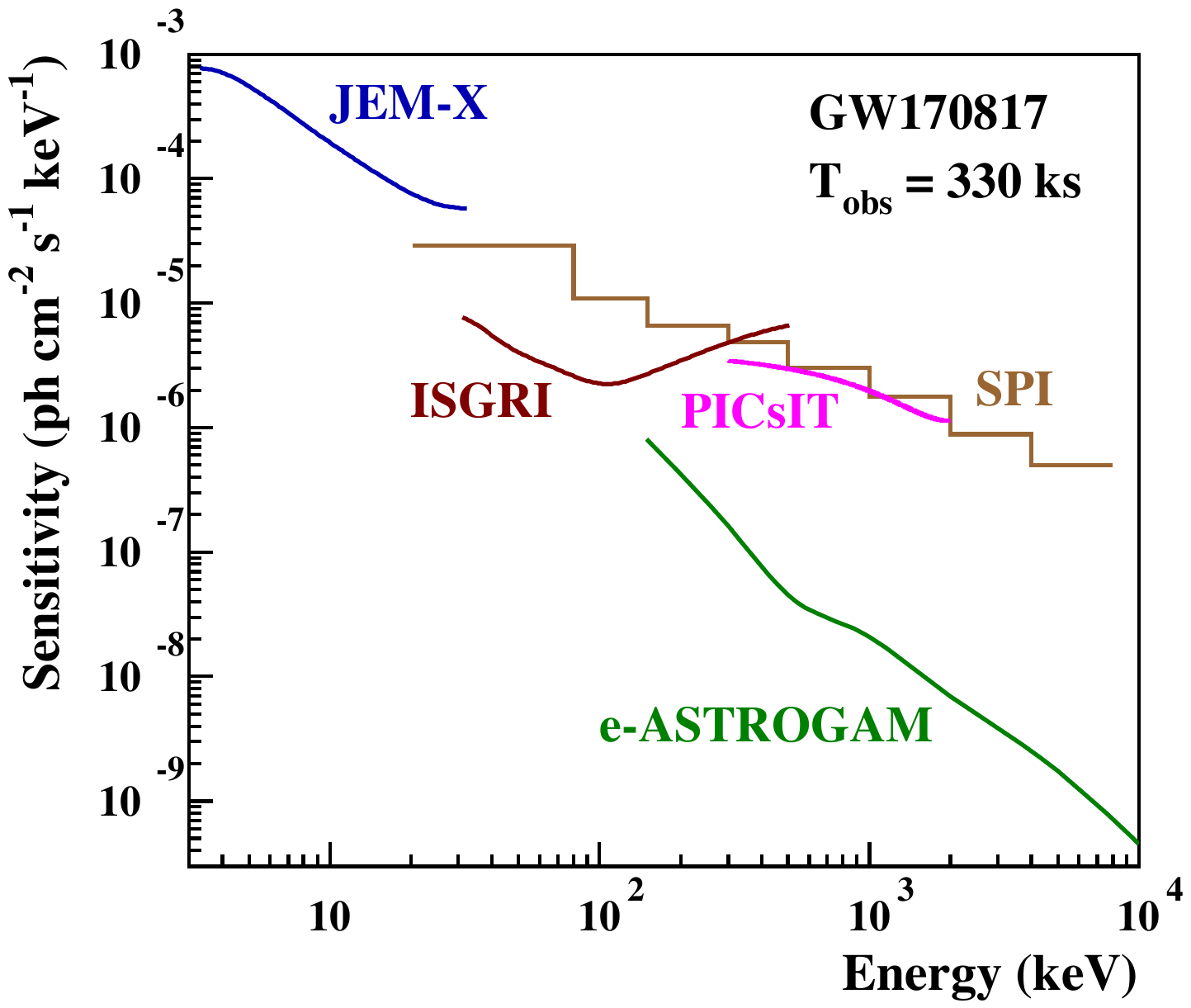}
\end{minipage}
\begin{minipage}{0.53\linewidth}
\includegraphics[scale=0.525]{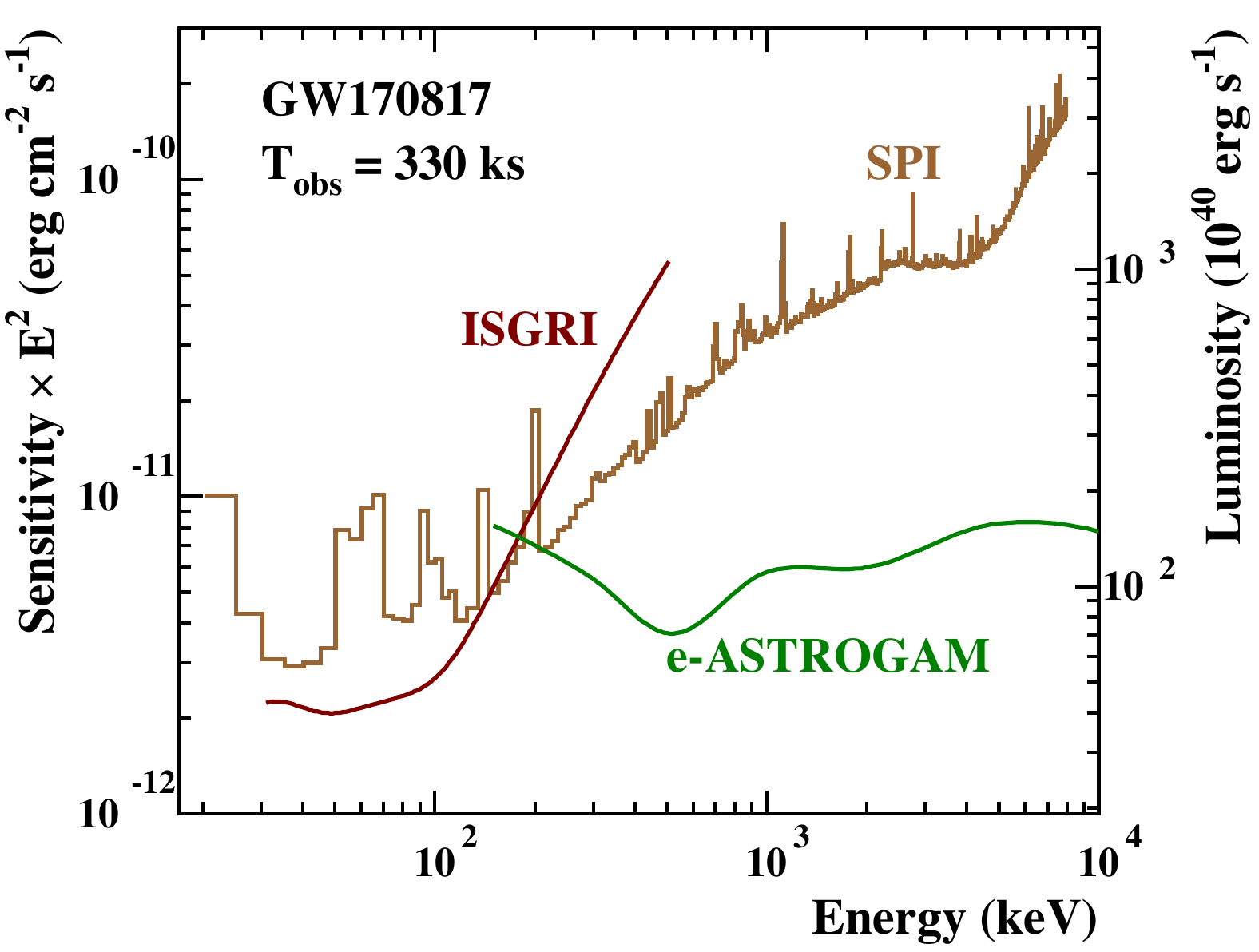}
\end{minipage}
\\
(a) \hspace{8.2cm} (b)
\end{tabular}
\end{center}
\vspace{-0.1cm}
\caption 
{ \label{fig:GW170817}
(a) Continuum and (b) narrow-line sensitivities reached in the {\it INTEGRAL} targeted follow-up observation of GW170817, compared to the corresponding sensitivities of e-ASTROGAM. All sensitivities are shown for a total exposure of 330~ks. The luminosity units of panel (b) assume a distance to the source of 40 Mpc. Adapted from Figs.~5 and 6 of Ref.~\cite{sav17}.} 
\end{figure} 

The follow-up of GW170817 conducted at optical, infrared and ultra-violet wavelengths revealed the presence of an EM counterpart with emission consistent with a kilonova powered by the radioactive decay of r-process nuclei synthesized in the merger outflows \cite{arc17,pia17,sma17}. \ea continuum and line sensitivities in the MeV range could allow to detect the radioactive decay gamma-rays from nearby kilonovae, thus directly probing the nucleosynthesis process thought to be responsible for the creation of approximately half the abundances of the nuclei heavier than iron.  
Figure \ref{fig:GW170817} illustrates the superior sensitivity of e-ASTROGAM (compared to {\it INTEGRAL}) to detect the continuum and nuclear line emissions  expected from the kilonova following a binary NS merger event like GW170817. The predicted gamma-ray line emission \cite{hot16} could be detected with e-ASTROGAM up to a maximum distance of $\sim 15$~Mpc. 

\ea would coincide with the third generation of ground-based interferometric GW observatories, such as the Einstein Telescope and the Cosmic Explorer. It could also be in operation at the same time as the Laser Interferometer Space Antenna (LISA), which will open GW observations to supermassive BHs expected to have magnetized circumbinary discs powering EM emission. Joint GW and gamma-ray detections by LISA and \ea could also come from binary systems of supermassive BHs, such as the BL Lac blazar PG 1553+113 \cite{tav18}.

\begin{figure}[t]
  \begin{center}
\includegraphics[scale=0.5]{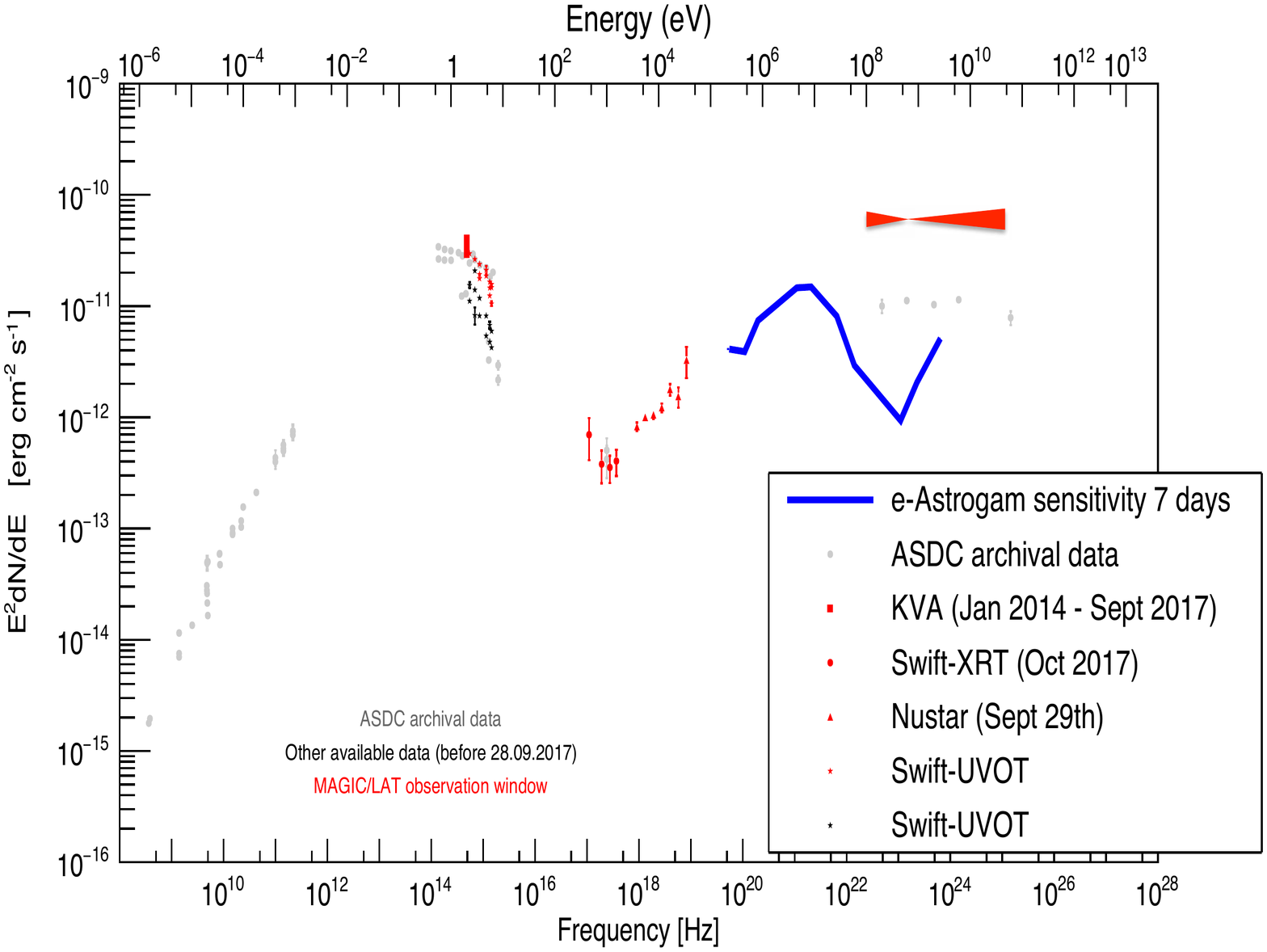}
  \end{center}
\caption{SED of the blazar TXS~0506+056. The red data points correspond to the flare observed by {\em Fermi} LAT (red butterfly) and MAGIC associated to the IceCube neutrino detection, while gray and black points show archival data. The e-ASTROGAM expected sensitivity (in blue) is calculated for an effective exposure of 7 days.}
\label{fig:TXS_SED}
\end{figure}

\subsection{e-ASTROGAM  and  neutrino astronomy}
\label{ssect:gammanu}

e-ASTROGAM's sensitivity is adequate to measure the spectral energy distribution (SED) of neutrino sources detectable by IceCube, as one can see from the recent detection of the association of a 300-TeV neutrino event \cite{icecubeatel} with an extended gamma-ray flare \cite{fermiatel,magicatel} of the Active Galactic Nucleus TXS0506+056 (Fig. \ref{fig:TXS_SED}).

Neutrinos are unique probes to study high-energy cosmic sources, since they are not absorbed by pair production via $\gamma\gamma$ interactions. Astrophysical high-energy neutrinos at TeV--PeV energies are generated \cite{deapim} by the decay of charged pions produced in inelastic photo-hadronic ($p\gamma$) or hadronuclear ($pp$) processes, involving protons $\sim20$ times more energetic than the resulting neutrinos. Photoproduction of neutrinos (and photons) via pion decay happens mainly via the $\Delta^+$ resonance slightly above kinematical threshold: $p\gamma \rightarrow \Delta^+ \rightarrow N \pi$.  
The energy of the proton has to be $E_p  \gtrsim$ 350 PeV/$\epsilon$, where $\epsilon$ is the target photon energy in eV. For UV  photons, as expected in AGN jets, this translates into $E_p \gtrsim$~10 PeV, i.e., above the knee: photoproduction of neutrinos on optical/UV photons is a likely indicator of UHECR acceleration. A simultaneous emission of hadronic gamma rays happens in both processes. An approximate relation holds {\em{at emission}} between the spectral production rates  of neutrinos and gamma rays in hadronic production: 
\[
E_\nu^2 \frac{dN_\nu(E_\nu)}{dE_\nu} \sim \frac{3K}{4}  E^2_\gamma \frac{dN_\gamma(E_\gamma)}{dE_\gamma} \]
with $K = 1/2(2)$ {{for}} \, $\gamma p \,(pp)$. Depending on the source optical depth, such photons are likely to further cascade, degrading their energy;  time and energy correlation between neutrinos and gamma rays  allows the measurement of the column density of the target.

\begin{figure}
\includegraphics[width=0.99\textwidth]{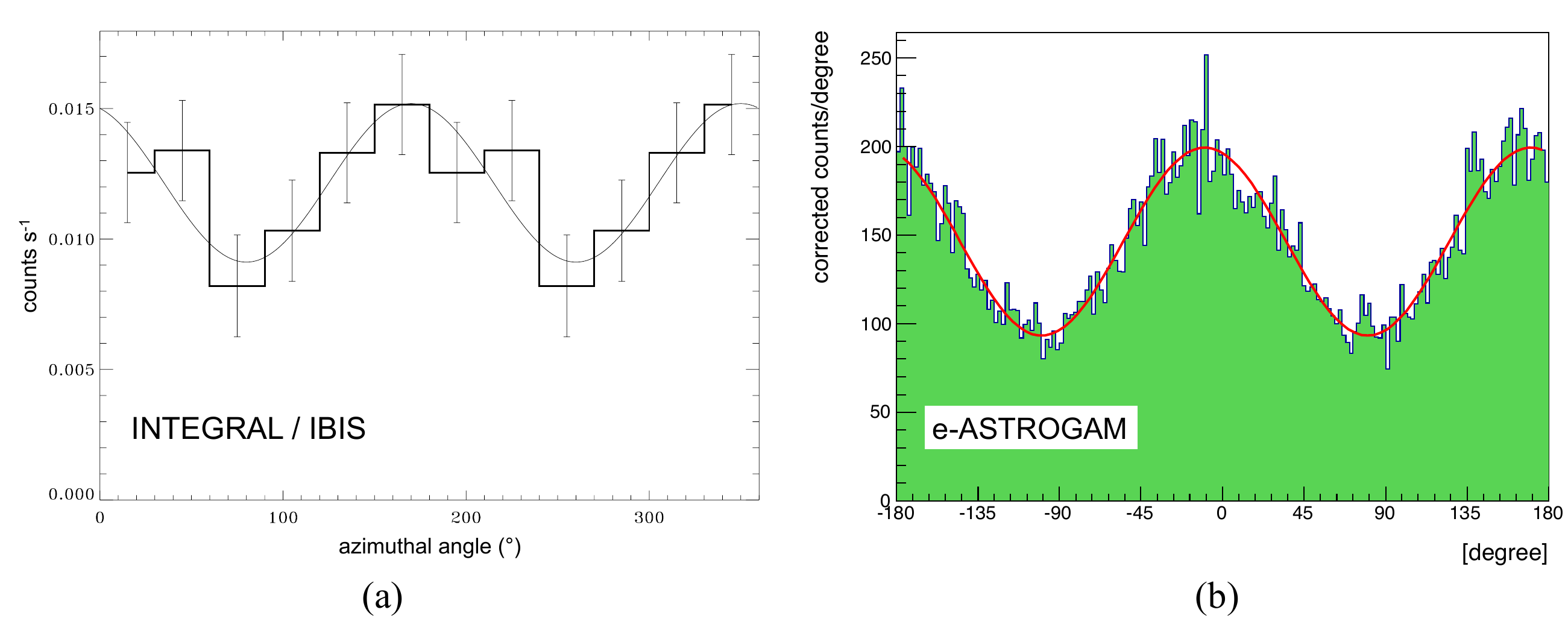}
\caption 
{ \label{fig:Crab_pol}
(a) Phase-averaged polarization diagram of the Crab pulsar and nebula obtained with {\it INTEGRAL}/IBIS in the 300 -- 450 keV energy band by summing 1.831~Ms of data taken between 2012 and 2014 (adapted from Ref.~\cite{mor16}). The measured polarization fraction and angle are ${\rm PF}=(98\pm37)$\% and ${\rm PA}=80^\circ \pm 12^\circ$, respectively. (b) Simulation of the e-ASTROGAM polarization signal in the same energy band for an effective duration of observation of the Crab pulsar and nebula 100 times shorter than that accumulated by {\it INTEGRAL}/IBIS (i.e. $T_{\rm obs} = 1.831 \times 10^4$~s $ \simeq 5$~h). The reconstructed polarization fraction and angle are ${\rm PF}=(98.0 \pm 2.4)$\% and ${\rm PA}=80.2^\circ \pm 0.7^\circ$.} 
\end{figure}

\section{Polarimetry with e-ASTROGAM}
\label{sect:pol}  % \label{} allows reference to this section

e-ASTROGAM will have a breakthrough capacity for gamma-ray polarimetry thanks to the fine 3D position resolution of both the Si Tracker and the Calorimeter, as well as the light mechanical structure of the Tracker, which is devoid of any heavy absorber in the detection volume. Fig.~\ref{fig:Crab_pol} provides an example of the expected quality of the e-ASTROGAM polarization data compared to currently available observations of the Crab pulsar and nebula with {\it INTEGRAL}. We see that e-ASTROGAM would accumulate a much higher statistics than {\it INTEGRAL}/IBIS in 100 times less exposure. Thus, with its excellent sensitivity and timing resolution, e-ASTROGAM will be able to provide a detailed gamma-ray view of the time-dependent polarization properties of the Crab nebula and pulsar, in particular during the periods of strong gamma-ray flaring activities possibly associated with a new type of particle acceleration \cite{abd11,tav11}. For example, a total exposure of 1~Ms would allow a {\it phase-resolved} polarimetric study of the Crab pulsar with a time resolution of only 20~$\mu$s, which is $1/1650$ of the pulsar rotation period.

At low energies (0.2 -- 2~MeV), e-ASTROGAM will achieve a Minimum Detectable Polarization (MDP) at the 99\% confidence level of 20\% for a 5~mCrab source after one year of effective exposure. With such a performance, e-ASTROGAM will be able to study the polarimetric properties of many pulsars, magnetars, gamma-ray binaries, and BH systems in the Milky Way. In the pair creation range, the measurement of polarization using the azimuthal orientation of the electron-positron plane is more challenging. But preliminary simulations show that a MDP$_{99}$ of $\approx 20$~--~40\%  in the energy range from 10 to 100 MeV is within reach for bright gamma-ray sources such as the Vela and Crab pulsars.  

Table~\ref{tab:polar} summarizes our conservative estimates of the number of sources for which good polarization data are expected with e-ASTROGAM. The cumulative number of GRBs to be detected as a function of MDP$_{99}$ was estimated from the simulated response of the gamma-ray instrument to linearly polarized GRBs at several off-axis angles in the range $[0^\circ;90^\circ]$. The GRB detection rate is expected to be $\sim 240$ per year, and a polarization fraction greater than $50$\% could be measured in $\sim 100$ GRBs per year. Such a high degree of linear polarization is consistent with the first GRB polarization results obtained by the Gamma-Ray Burst Polarimeter (GAP) aboard the IKAROS solar sail mission \cite{yon12}, and by {\it INTEGRAL} \cite{mcg07,got09}. Polarization information in GRBs can place strong new observational constraints on the nature of the ultrarelativistic outflow, its geometry and magnetization, as well as on the high-energy radiation mechanisms; measurements of GRB polarization also have the potential to provide constraining limits on violation of Lorentz invariance \cite{lau11}. 

Polarization measurements of blazars can provide crucial insight to the geometry of the emitting regions and allow to discriminate among various processes proposed as emission mechanisms. In particular, such measurements have the potential to identify the presence of hadrons in jets of BL Lac blazars, as hadronic models predict a much higher degree of polarization than leptonic models \cite{zha13}. e-ASTROGAM is expected to provide good polarization data for at least the 15 blazars already detected by COMPTEL \cite{col06} and SPI \cite{bou08}, as well as for the two very bright BL Lac objects PKS~2155-304 and Mkn~501. However, with its unprecedented sensitivity in the 1~--~100~MeV energy range, e-ASTROGAM should detect more than $\sim 350$ blazars in three years (Table~\ref{tab:nevents}), mostly FSRQs, and good polarization data could be obtained for a significant fraction of them. 

e-ASTROGAM is also expected to detect the gamma-ray polarization signal of a handful of Seyfert galaxies, and the measured polarization fraction and angle will provide important information on the geometry of the Comptonizing medium in these active galactic nuclei (AGN). Good bright examples are the Seyfert 1 galaxies NGC~415 and IC~4329a, and the Seyfert~2 galaxies NGC~4945 and Cen~A, all of them detected above 200 keV by {\it INTEGRAL} SPI \cite{bou08,bur14}, as well as the archetypal Compton thick AGN NGC 1068 detected by {\it Fermi}-LAT \cite{len10}.

e-ASTROGAM will be able to study the polarimetric properties of the 18 currently known, rotation-powered, soft gamma-ray pulsars \cite{kui15}, which are thought to be significantly younger and more energetic than the pulsars detected in the GeV domain by {\it Fermi}-LAT \cite{abd13}. Detailed phase-resolved polarimetry of the pulsed emission will be possible for a handful of bright MeV pulsars, such as PSR B0531+21 (Crab, see above), PSR B1509-58 (MSH 15-52) and PSR J1930-1852 (G54.1+0.3). 
The polarimetric measurements with e-ASTROGAM will shed new light on the processes of particle acceleration and pair cascades in the pulsar magnetosphere, as well as on the geometry and structure of the magnetic field in these systems. 

At variance with ordinary radio-pulsars, magnetars are ultra-magnetized NS ($B \approx 10^{13} - 10^{15}$~G) powered by their own magnetic energy \cite{tur15}. e-ASTROGAM is expected to detect the polarization of both steady and flaring emissions from magnetars. Polarization measurements of the persistent soft gamma-ray emission of four magnetars (see Table~\ref{tab:polar}) will reveal the reprocessing of thermal photons emitted by the NS surface through resonant Compton scattering in the twisted magnetosphere. e-ASTROGAM should also provide invaluable information on NS physics by measuring the polarization of strong bursts from Soft Gamma Repeaters (SGR). 

\begin{table}[t!]
\begin{center}
\begin{tabular}{| l | c | c |}\hline
Type & Nb sources & Comments\\ \hline
GRBs         & $\sim 300$  & Minimum detectable polarization of $10$\%, $20$\%, and \\
             &             & $50$\% in $\sim 16$, 40, and 100 GRBs per year, respectively\\ 
Blazars      & $>17$       & 13 FSRQ and 4 BL Lacs (PKS 0716+714, PKS 2155-304,\\
             &             & Mkn 421 and Mkn 501)\\
Other AGN    & $>6$        & Bright radio-quiet quasars and Seyfert galaxies: \\
             &             & NGC 4151, IC4329a, NGC 4945, NGC~1068, 3C~120, Cen A \\
Pulsars \& PWNe & $> 18$  & 18 currently known soft gamma-ray pulsars \cite{kui15}\\ 
Magnetars    & $> 4   $    & Persistent gamma-ray emission of 1RXS J1708-4009, \\ 
             &             & 4U 0142+61, 1E 1841-045, and SGR 1806-20\\ 
$\gamma$-ray binaries & $> 6 $& Currently known $\gamma$-ray binaries: PSR B1259-63, \\ 
             &             & LS I +61 303, LS 5039, HESS J0632+057, \\ 
             &             & 1FGL J1018.6-5856, and CXOU J053600.0-673507 \\ 
Microquasars & $> 9  $     & Cyg X-1, Cyg X-3, GRS 1915+105, 1E1740.7-2942,  SS 433, \\ 
             &             & GRS 1758-258, GX 339-4, XTE J1550-56, GRO J1655-40\\ 
\hline
\end{tabular}
\end{center}
\caption{Estimated number of sources of various classes whose polarization properties will be measured with e-ASTROGAM in three years of nominal mission duration. \label{tab:polar}}
\end{table}

Gamma-ray binaries are known to be powerful MeV emitters \cite{dub13} and in the two scenarios describing the particle acceleration in these objects --~(i) jets of a 
microquasar powered by accretion from a massive star and (ii) shocks between the relativistic wind of a young non-accreting pulsar and the wind of the massive stellar companion~-- the resulting nonthermal gamma-ray emission is expected to be polarized. e-ASTROGAM has the capability to measure the polarization degree of the six currently known gamma-ray binaries (Table~\ref{tab:polar}). 

Gamma-ray polarimetry of microquasars with e-ASTROGAM is also very promising. Simulated polarigrammes for e-ASTROGAM observations of Cyg X-1 in the ``low'' hard state show that the mission will be able to provide a detailed spectral and time-resolved gamma-ray view of the polarization properties of this system on a timescale of $10^4$~s, thus answering questions regarding the different emitting media (Comptonized corona vs. synchrotron-self Compton jets), while providing important clues on the composition, energetics and magnetic field of the jet. The excellent sensitivity of e-ASTROGAM in the MeV range will also give access to detailed studies of other, fainter, microquasars (e.g., Cyg X-3, GRS 1915+105, 1E1740.7-2942, see Table~\ref{tab:polar}), whose polarized gamma-ray emission is not currently detectable.

\section{Summary}

e-ASTROGAM is a concept for a gamma-ray space observatory that can revolutionize the astronomy of medium/high-energy gamma rays by increasing the number of known sources in this field by more than an order of magnitude and providing polarization information for many of these sources -- thousands of sources are expected to be detected during the first 3 years of operations. Furthermore, the proposed wide-field gamma-ray observatory will play a major role in the development of time-domain astronomy, and provide valuable information for the localization and identification of gravitational wave sources. 

The instrument is based on an innovative design, which minimizes any passive material in the detector volume. The instrument performance has been assessed through detailed simulations using state-of-the-art  tools and the results fully meet the scientific requirements of the proposed mission. 

e-ASTROGAM will operate as an observatory open to the international community. The gamma-ray observatory will be complementary to ground and space instruments, and multifrequency observation programs will be very important for the success of the mission. In particular, e-ASTROGAM will be essential  for investigations jointly done with radio (VLA, VLBI, ALMA, SKA), optical (JWST, E-ELT and other ground telescopes), X-ray (Athena) and TeV instruments (CTA, HAWC, LHAASO and other ground-based detectors). Special emphasis will be given to fast reaction to transients and rapid communication of alerts. New astronomy windows of opportunity (sources of gravitational waves, neutrinos, UHECRs) will be fully and uniquely explored.

\acknowledgments % equivalent to \section*{ACKNOWLEDGMENTS}       
 
The research leading to these results has received funding from the European Union's Horizon 2020 Programme under the AHEAD project (grant agreement n. 654215).

% References
\bibliography{report} % bibliography data in report.bib
\bibliographystyle{spiebib} % makes bibtex use spiebib.bst

\end{document}